\documentclass{an}

\usepackage{graphicx}
\begin{document}
\headnote{AN 322 (2001) 4, 000--000}

\title{A new concept and a preliminary design for a high resolution (HR)
and very-high resolution (VHR) spectrograph for the LBT}

\author{F.~M.~Zerbi\inst{1} \and P.~Span\`o\inst{2}
\and P.~Conconi\inst{1} \and E.~Molinari\inst{1} \and R.~%
Mazzoleni\inst{1} \and R.~Pallavicini\inst{2} \and K.G.~Strassmeier\inst{3}} \institute{ INAF -
Osservatorio Astronomico di Brera, Via Bianchi 46, I-23807 Merate
(LC), Italy \and INAF - Osservatorio Astronomico di Palermo,
Piazza del Parlamento 1, I-90134 Palermo, Italy \and
Astrophysikalisches Institut Potsdam (AIP), An der Sternwarte 16,
D-14458 Potsdam, Germany }

\date{Received {\it -----------}; accepted {\it ------------}}

\abstract{A way to fully exploit the large collecting area of
modern 8--10m class telescopes is high resolution spectroscopy.
Many astrophysical problems from planetary science to cosmology
benefit from spectroscopic observations at the highest resolution
currently achievable and would benefit from even higher resolutions.
Indeed in the era of  8--10m class telescopes no longer the telescope
collecting area but the size of the beam --- which is related to the
maximum size in which reflection gratings are manufactured --- is
what mainly limits the resolution. A resolution-slit product
$R\varphi \simeq 40,000$ is the maximum currently provided by a beam
of 20 cm illuminating the largest grating mosaics.
We present a
conceptual design for a spectrograph with $R\varphi \simeq 80,000$, i.e.
twice as large as that of existing instruments. Examples of the
possible exploitation of such a high $R\varphi$ value, including
spectropolarimetry and very high resolution ($R\sim 300,000$), 
are discussed in detail. The new concept is illustrated
through the specific case of a high resolution
spectropolarimeter for the Large Binocular Telescope.
 }

\maketitle

\section{Introduction}

The construction of 8--10m class telescopes has
brought new opportunities for astrophysical research using high
resolution spectroscopy. Over the last few years a number of
spectrographs have been designed and manufactured for telescopes
of this class, most of them with large optics, mosaic gratings,
fast focal ratio cameras and large area CCD detectors (e.g.
Pilachowski et al. 1995). Large telescopes allow us to extend high
resolution spectroscopy to weaker sources, but a number of
problems arise from the related larger beam diameters. Such
problems are related to the demand for higher stability, good
velocity accuracy and careful attention to the instrumental profile.

The above demand has been fulfilled via new spectrograph concepts
such as \emph{white pupil} design, immersed gratings, active pupil
correction and larger refractive or catadioptric cameras. The
performances of such spectrographs are always a compromise between
resolution, throughput, and spectral coverage.

The dispersing element adopted in all high resolution
spectrographs studied or built so far is the \emph{echelle}
reflection grating. From diffraction theory (e.g. Schroeder, 2000) we
know that the 
resolution of a conventional echelle spectrograph in Littrow
configuration is given by the \emph{resolution-slit product}:
\begin{equation}
  R \varphi = 2 (d/D) \tan\delta
 \end{equation}
where $R$ is the resolution, $\varphi$ is the angular slit width on
the sky (as seen by the telescope), $d$ is the beam diameter at
the collimator, $D$ is the telescope diameter and $\delta$ is the
blaze angle, i.e., the angle between the facet and the grating
normals.

Modern high resolution spectrographs for 8--10m class telescopes
have a maximum resolution-slit product of about 40,000 arcsec.
Such a number comes from state-of-the-art technology in
optical component manufacturing. The straightforward way to
increase the  \emph{resolution-slit product}, for a telescope of a
given aperture $D$, is to increase the blaze
angle $\delta$ and/or the beam diameter $d$. R4 echelles, 
corresponding to $\delta=76^{\circ}$, are the most tilted gratings
currently manufactured. Construction difficulties prevent larger
blaze angles and large beam diameters as well, not to talk about costs. 
Indeed the beam diameter
is limited by the maximum dimensions of the grooves area in
currently available ruling engines, i.e. about 20 cm. Other ways
have to be explored to further increase resolution.

In this paper we propose a new optical concept based on the use of
two R4 echelle reflection gratings allowing us to double the currently
achievable $R\varphi$ values. Cross-dispersion by means of high
efficiency volume phase holographic (VPH) gratings is used to
compensate for the reduced efficiency of using two echelle gratings. 
This concept will be illustrated
in the case of a high-resolution (HR) and very-high resolution
(VHR) spectrograph proposed for the Large Binocular Telescope (LBT).

\section{Scientific motivations for HR and VHR spectroscopy}

High resolution spectroscopy (i.e. spectroscopy at resolving powers of
R$\simeq$40,000 or higher) is a fundamental tool to derive physical information
about cosmic sources, both galactic and extragalactic. The optical
spectra of astrophysical objects can provide, if observed at high 
resolution and high S/N, detailed information on many important physical
quantities such as chemical abundances and their ratios, radial and
rotations velocities, convective motions, pulsations, mass losses, 
magnetic fields, and many others (see the proceedings of the
Workshop on \emph{``High-Resolution Spectroscopy with Very Large Telescopes''},
Kraft 1995, and reference theirein).
Until recently, the biggest limitation to high
resolution spectroscopy has been the small aperture of the available
telescopes, which limited HR spectroscopy mostly to stellar physics and
relatively bright objects. With the advent of 8--10m class telescopes,
equipped with HR cross-dispersed echelle spectrographs and sensitive CCD
detectors, it is now possible to extend the realm of HR spectroscopy to
much fainter stars and to extragalactic sources as well. Moreover, the
large aperture of the current and future telescopes allows a
further stepforward in resolution, opening up the possibility of very
high resolution (at resolving power of several 100,000)
for objects well beyond the brightest and closest stars.

HR spectrographs working at resolutions of 40,000--100,000 on 8--10m
class telescopes (as well as on 4m class telescopes for brighter objects) have
already demonstrated the power of HR spectroscopy in many different
research areas. Stellar objects remain the most easily accessible targets, 
but absorption lines in QSOs and distant galaxies have also become observable,
at least for the brightest of these objects,
thus providing information on the intergalactic medium and absorbing
material along the line of sight (e.g. Shull 1995).
High resolution is needed to resolve the complex velocity fields of the
intervening gas, while the need for high throughput is dictated by the 
faintness of most extragalactic objects.

HR is also crucial to resolve the velocity structure of interstellar 
medium clouds in our Galaxy (e.g. Kennicutt et al. 1995),
as well as to determine the $^6$Li/$^7$Li isotope ratio in the ISM,
a fundamental parameter to investigate the production of Li 
by cosmic rays during galactic evolution. With large aperture telescopes, 
it is now possible to investigate the ISM throughout the Galaxy and in many 
different directions.

Abundance determinations in stars of different populations remain a 
fundamental tool to study stellar evolution and the formation of the 
galactic disk, of the bulge and of the halo (e.g. Sneden et al. 1995). 
Although abundance determinations can be made in some cases through
equivalent widths rather than line profiles, high resolution is required 
in many cases to resolve blends and for accurate matching of the observed 
line profiles with synthetic ones. Chemical abundances and their
ratios, as well as isotopic ratios, provide information on internal
mixing in stars and put stringent constraints on models of stellar
structure and 
evolution. Observations of open and globular clusters, 
i.e. of homogeneous samples of stars with the same age and chemical 
composition, allow tracing galactic evolution for different 
stellar populations and in different parts of the Galaxy. Determinations 
of the Li and Be abundance in stars allow investigating stellar 
interior mixing processes as well as primordial nucleosynthesis
and Li enrichment throughout the history of our Galaxy (Pinsonneault
1997).

The determination of accurate line profiles in the spectra of stars
allow studying the velocity fields in stellar atmospheres  
produced by the interaction of convection and rotation and/or stellar
oscillation. Line profiles (e.g. Demarque 1995), asymmetries and
shifts and line bisectors observed at high resolution and high S/N in
late-type stars allow us to derive information on convective motions
generated in subphotospheric convective zones (Gray 1992). 
They also provide information on winds and mass losses in hot stars and
in late-type giants and supergiants. Accurate line profiles also allow us 
to measure magnetic fields (through the Zeeman effect) and to infer
the presence of magnetic surface structures line spots and  
plage (through Doppler imaging techniques, e.g. Vogt \& Penrod 1983). 

Finally, the determination of precise radial velocities by means 
of HR spectroscopy allow us to search for planets orbiting nearby stars, 
as well to determine stellar pulsations and oscillations (e.g. Queloz
2001).  The field of stellar asteroseismology of solar type stars 
is still in its infancy but 
it is expected to grow enourmously with the advent of high stability  
HR spectroscopic facilities at large telescopes.  However
asteroseismology of other classes of objects such as white dwarfs,
$\beta$ Cephei or $\delta$ Scuti stars is an established science
and benefits of well developed methods, procedures and
interpretative models based on HR spectroscopy data (e.g. Zerbi
2000). The existence of 
planetary size bodies orbiting nearby stars has been clearly 
demonstrated on the basis of accurate radial velocity measurements, 
but much remain to be done for understanding the formation of planets 
and planetary systems other than our own.

To study velocity fields in stellar atmospheres as well as to study
isotope ratios and velocity fields in the ISM, spectral resolutions
around 100,000 are often insufficient. Resolutions at least a factor 
2 or 3 higher are desirable together with high throughput of the 
telescope-spectrograph combination. The
large aperture of 8-10m class telescopes offers this possibility
for the first time. A VHR spectrograph (R$\simeq$300,000) on an 8m class 
telescope would be a unique instrument, not yet available at any large
telescope in the world. From line profiles, asymmetries and shifts
measured by such an instrument, we could address the physical properties
of convection, rotation, magnetism, stellar activity and stellar
seismology at a level of detail so far possible only in the case 
of the Sun. Fields such as asteroseismology of solar-type stars,
detection of extra solar planets and the study of the interstellar
medium are all expected to greatly benefit from VHR spectroscopy.

The possibility of analysing the light of celestial sources in their
different polarization components offers the further advantage of
studying the central role of magnetic fields in astrophysics. Under this
respect, coupling a HR spectrograph with a polarimetric unit is the only
way to understand the structure and dynamics of surface magnetic fields
for a variey of objects, ranging from late-type stars to young stellar
objects, from accretion sources in close binaries to QSOs and AGNs 
(Strassmeier et al.~2001). The spectrograph we will discuss in the following
sections offer these various capabilities, including a range of possible
resolutions (from HR to VHR) as well polarimetric capabilities.

\section{High resolution spectrographs and their limitations}

\subsection{Overview of existing instruments}

Five high resolution cross-dispersed echelle spectrographs are
currently operative or under construction at large telescopes
(diameter 8--10m): HIRES at Keck, UVES at VLT, HRS at HET, HDS at
Subaru, HROS at Gemini South. They provide $R\varphi$ products of 
$\sim$40,000 and maximum resolutions, with lower efficiency, of 
$\sim$100,000 or slightly higher using image slicers. 
Their characteristics are summarized
in Table~\ref{tab:1}.

\begin{table*}
 \small
  \caption{High resolution spectrographs at 8--10m class
  telescopes. The last column shows the performances of the
 spectrograph we are proposing for the LBT; see later.}
 \centerline{
 \begin{tabular*}{\textwidth}{@{\extracolsep{\fill}}|l|lllll|l|}  \hline
   Spectrograph & HIRES & UVES & HRS & HROS & HDS & PEPSI \\ \hline
    Telescope & Keck & VLT & HET & Gemini & Subaru & LBT \\
    Diameter (m) & 10 & 8.2 & 9.2 & 8.1 & 8.2 & 2x8.4 \\
    Area (m$^2$) & 76 & 51.2& 77.6& 50.4& 51.3& 109.7 \\   \hline
  $R\varphi$ (arcsec) & 39,000 & 40,000 & 30,000 & 28,500 & 36,000 & 78,000 \\
  Max. R & 67,000 & 115,000 & 120,000 & 75,000 & 160,000 & 320,000\\
  Focal station & Nasmyth & Nasmyth & fiber fed & Cassegrain & Nasmyth
 & fiber fed\\
  $\lambda\lambda$ ($\mu$m) & 0.3--1.1 & 0.3--1.1 & 0.4--1.1 &
  0.3--1.0 & 0.3--1.0 & 0.5--1.0 \\
  Collimator & f/13.7 & f/10 & f/10 & f/16 & f/12.5 & f/10 \\
  Beam diameter (mm) & 305 & 200 & 200 & 160 & 272 & 160\\
  Echelle(s) & 1x3 mosaic & 1x2 mosaic & 1x2 mosaic & immersed & 1x2
  mosaic & 2 1x2 mosaics \\
  Blaze angle & R2.8 & R4 & R4 & R2 & R2.8 & R4\\
  Cross-disperser & grating & gratings & gratings & prisms & gratings
 & VPH gratings \\
  Detector & CCD & CCD mosaic & CCD mosaic & CCD mosaic & CCD mosaic &
  CCD mosaic\\
  Spectral format & 2Kx2K & 4Kx4K  & 4Kx4K & 2Kx4.6K & 2Kx4K & 8Kx4K \\
  Pixel ($\mu$m) & 24 & 15 & 15 & 13.5 & 13.5 & 15\\
  Limiting magnitude$^a$ & 19.7 & 19.4 & 19.4$^b$ & NA$^c$ & 18.6 &
  20.2  \\ \hline
  \multicolumn{7}{l}{$^a$ at $R\simeq$60,000 in 1 hour exposure
 with S/N=10 per resolution element, as provided by the ETCs of 
 each instrument,}\\
  \multicolumn{7}{l}{where available}\\ 
  \multicolumn{7}{l}{$^b$  V=19.4 was estimated by Tull (1998); however a lower
 value is showed in the web page}\\
  \multicolumn{7}{l}{({\tt http://rhea.as.utexas.edu/HET\_hrs.html})}\\
  \multicolumn{7}{l}{$^c$ NA, not available.} \\
  \end{tabular*}}
  \label{tab:1}
\end{table*}
\normalsize

\begin{figure}
   \resizebox{\hsize}{!}
   {\includegraphics{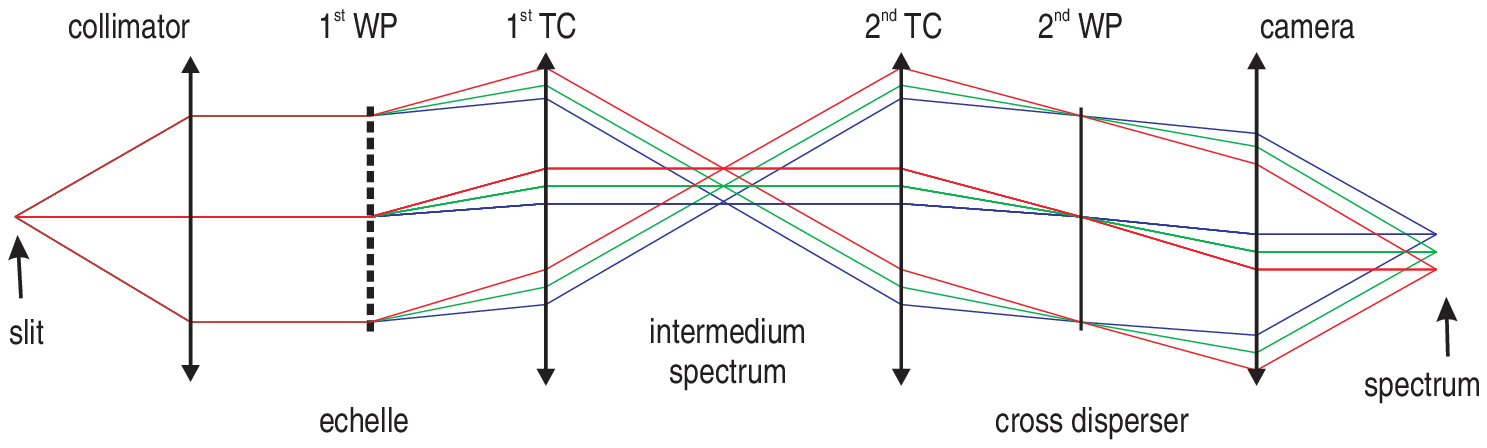}}
  \caption{\textbf{Double white pupil: schematic design.}
  The echelle is located on
  the first white pupil (1$^{st}$ WP). Two transfer collimators (TC)
  reimage a second white pupil (2$^{nd}$ WP) where the cross
  disperser is located. This pupil corresponds to the entrance pupil of
  the camera.}
  \label{fig:white1}
\end{figure}

Two of the instruments in Table~\ref{tab:1} (UVES at VLT and
HRS at HET) are based on the
\emph{white pupil} design. The \emph{white pupil} concept has been
introduced by Baranne (1972) and elaborated by Delabre for UVES
(Dekker et al.~2000). This configuration eliminates vignetting and
aberrations by re-imaging the pupil at the echelle onto the camera
pupil, where the cross disperser is located (see
Figure~\ref{fig:white1}). With few added optical elements, providing
 only a few percent of light-loss,
the vignetting typical of traditional designs is removed and the
image quality and overall luminosity substantially enhanced.
\begin{figure}
   \resizebox{\hsize}{!}
   {\includegraphics{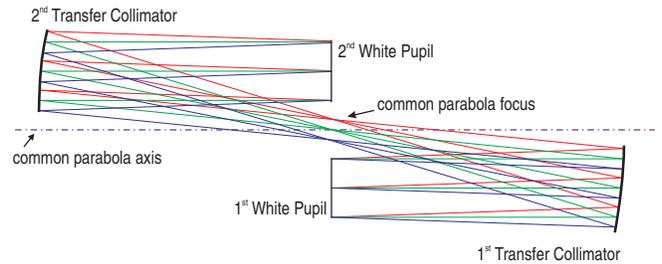}}
   \caption{\textbf{Double white pupil: an example.}
   This optical system consists of two identical off-axis parabolae
   with the same axis and the same focus, that reimage the first
   white pupil onto the second one with magnification 1.}
   \label{fig:white}
\end{figure} 

UVES (UV--Visual Echelle Spectrograph) for the ESO Very Large
Telescope is based on a 1x2 mosaic R4 echelle that provides
$R\varphi \simeq 40,000$ with a beam size of 200 mm.
The steep angle of the R4 grating requires minimizing the angle
between camera and collimator to preserve the efficiency of the
grating. UVES is mounted at an f/15 Nasmyth focus of unit UT2
(Kueyen)
 of the VLT. It has two arms, a red one (420--1100 nm) and a blue one
(300-500 nm) to optimize efficiency. Many pre-slit optical systems
are common to the two arms (image derotator, ADC, depolarizer,
dichroic, image slicers, I2 cell). The two arms have the
same optical design: a main off-axis parabolic collimator is used
in double pass once to collimate the beam on the grating and a
second time to reform an intermediate spectrum very near the
slit location. A second identical collimator cancels the off-axis
aberration of the first one and casts a white pupil onto the cross
disperser.

Since the camera aperture in UVES is relatively small, there is no
need for very fast focal ratios for the cameras: f/1.8 for the blue
one and f/2.5 for the red one. $R\varphi=41,400$ (38,700) in the
blue (red) domain allows a maximum resolution of 80,000 (110,000) with a
0.4 (0.3) arcsec slit. UVES is equipped with a 2Kx4K
EEV CCD in the blue, and a mosaic of two different (EEV and MIT/LL
2Kx2K) CCDs in the red. With a S/N=10 \emph{per wavelength bin}, for a 1 hr
exposure, seeing = 0.7 arcsec, limiting magnitude at R=60,000 is
V=19.0 (red) and U=17.5 (blue). Using the iodine
cell the velocity accuracy is 47 m/s (35 m/s) in the blue (red)
domain. A performance comparison in terms of efficiency, spectral
coverage and resolution indicates that UVES is the most advanced
spectrograph available at present.

Another high resolution white pupil spectrograph is the High
Resolution Spectrograph (HRS, Tull 1998) at the 9m Hobby-Eberly
Telescope, that is essentially an arm of the UVES design. It is fed
from the
f/1.8 primary focus with fibers and it is mounted on an
optical bench, allowing high stability. It has a 200 mm beam
diameter and an off-axis parabolic collimator in white pupil
configuration. Its R4 1x2 mosaic echelle is identical to the UVES
red one. Its $R\varphi=30,000$ should allow reaching a limiting magnitude
V=19.4 at R=60,000, with a S/N=10 in 1 hr exposure.
However actual measurements (http://rhea.as.utexas.edu/HET\_hrs.html)
 show that the limiting magnitude is 2.5 mag lower.

HIRES (HIgh REsolution Spectrograph, Vogt et al. 1994) is
operative since 1993 at the Keck I telescope. In HIRES the
light enters the slit from the Keck f/13.7 Nasmyth focus, is
collimated by a spherical reflective collimator to form a 305 mm
beam that is directed to a 1x3 320x1270 mm$^2$ mosaic R2.8
echelle, mounted on a granite substrate. The beam is then
cross-dispersed by a low-order grating and fed into an f/1.0
catadioptric camera with 0.75 m of clear aperture that provides
excellent image quality from 0.31 to 1.1 $\mu$m over the full
6.7$^{\circ}$ camera field of view. In spite of its
$R\varphi=39,000$, maximum resolution is limited to 67,000 by the 2Kx2K 24
$\mu$m pixel CCD. For a S/N ratio of 10, a 1 hr exposure at
R=60,000 gives V=19.7.

HDS, the High Dispersion Spectrograph for the Subaru Telescope
(Noguchi et al. 1998) is very similar to HIRES, with a slightly
higher image quality and a smaller pixel size, ensuring better
spectral resolution. This instrument has entered in operation in April
2001 and the limiting magnitude is given as V=18.6 in 1 hour at S/N=10.

HROS, the High Resolution Optical
Spectrograph (D'Arrigo et al. 2000) for Gemini South, makes use
of an \emph{immersed} echelle grating (Szumski \& Walker 1999) in
order to decrease overall size and weight, since the instrument
will be mounted at the
Cassegrain focus through a pre-slit prism focal modifier. The R2 echelle is
immersed into a high refraction index medium. If we call $n$ such an
index, the resolution of the immersed grating is $n$ times higher
than conventional reflection gratings. With a $R\varphi=28500$,
HROS is capable of a maximum resolution of 75,000.
However, the instrument design is not yet finalised since stability
problems at the Cassegrain focus may eventually lead to modifications of
the current design. 

\subsection{Limits of existing configurations}

As stated in the introduction, existing high resolution
spectrographs are limited by the $R\varphi$ they can provide, i.e.
about 40,000 for an 8m class telescope. For a given $R\varphi$  and a
given echelle grating the
only way to increase the resolution is to narrow the entrance slit
of the spectrograph. A narrower slit however reduces the
instrument throughput and requires an adequate sampling.
For instance, an instrument with an $R\varphi=40,000$ can be
operated at $R=200,000$ provided a slit of $\varphi=0.2$
arcsec is used. The problem is that
the slit dimension prevents most of the source light to enter the
spectrograph, even in good seeing conditions.

It is possible to overcome this problem by using either image slicers
or adaptive optics (AO). Image slicers are the most widely used devices
to allow spectrographs to be operated at high resolution with
acceptable throughput. With the use of an image slicer the
resolution in the spectrographs reviewed above can be
pushed up to about $R \simeq 120,000$ (160,000 for HDS, see Table
\ref{tab:1}). The use of adaptive optics to increase the slit
throughput is instead the object of recent studies (Wiedemann et al.\
2000, Pallavicini~2001) and has found so far only limited 
practical application in the optical (Ge et al.\ 1999). We will come
back to this possibility later on.

Since the dimensions of the CCD detector are limited, increasing the
resolution of a spectrograph by narrowing the entrance slit also reduces
the available spectral coverage in a single exposure.
In the example used above, i.e. a spectrograph
with $R\varphi=40,000$ operated at $R=200,000$, the projected slit image
will cover a $\delta\lambda=0.004$ nm at 800 nm, and will have to be
sampled with at least 2.5 pixels.
Assuming a CCD with 15 $\mu$m pixels, we obtain a reciprocal
dispersion of 0.11 nm/mm. For a 4Kx4K CCD (typical of
existing spectrographs), one order will cover 5.5 nm at 800 nm.
If we want to cover the range between
500 and 1000 nm (\footnote{this will be the spectral range
we are interested in for the proposed instrument at the LBT.})
we will need (approximately) 100 orders.  In
order to accomodate 100 orders on the CCD, assuming an order
separation of the same size of the spectrum, the slit height is
limited to 1/200 of the CCD height, i.e., 
each spectrum has a maximum allowed height of 307 $\mu$m
corresponding to about 1.4 arcsec at the slit location. Such a
figure might not be a problem for a fiber-fed spectrograph where
the core radius of the fiber corresponds to the slit height but it
is certainly a problem for direct-feed spectrographs or for instruments
equipped with image slicers. In order to cover 500-1000 nm at higher
resolutions we are forced to either increase the CCD dimensions, or to
reduce the available spectral range in a single exposure.

Both the spectral coverage and throughput problems would be alleviated by
increasing the $R\varphi$ product. For example, a
$R\varphi \simeq 80,000$ spectrograph could be operated at $R=200,000$
with a 0.4 arcsec entrance slit (instead of 0.2) with an evident gain in
throughput. The portion of spectrum covered by each order remains
unchanged since the free spectral range is the same. However, the
height of the spectrum in each order and the inter-order
separation is decreased allowing to optimize the CCD filling. In
the case, for example, of a 10 arcsec slit-height, in the
$R\varphi=40,000$ configuration the spectrum height is 125 pixels
while in the $R\varphi=80,000$ setup the height is reduced to one
half of it, i.e. to 62.5 pixels, thus allowing to cover a spectral
range twice as large in a single exposure.

\section{A new concept}

A way to increase $R\varphi$ without increasing the size of
the  beam and/or the grating blaze angle
is represented by the use of 2 echelle gratings mounted ``in
series'', in order to increase angular dispersion. Such an idea
has been already used in the MEGA monochromator (Engman \&
Lindblom 1984). MEGA, built with $R$ about 2,000,000 using four
echelles, was not conceived for astronomical use and hence was not
optimized for efficiency. The concept however is suitable to be
optimized for astronomical use and to be combined with the
white-pupil design taking advantage of the optical quality given
by such configuration. In the following section we present a
triple pupil configuration accommodating two R4 echelle gratings and
volume phase holographic (VPH) gratings as cross-dispersers, capable of
providing a $R\varphi$ of the order of 80,000 with a beam size
similar to that of UVES, i.e. 20 cm.

We present a preliminary study of the above concept for 
application to the Large Binocular Telescope (LBT). This telescope,
for which no high resolution optical spectrograph has yet been selected,
will be the largest single telescope in the coming
years. The unprecedented throughput of LBT will allow us to extend
high-resolution (HR) spectroscopy to weaker sources and to perform
very-high-resolution (VHR) spectroscopy of a significant number of
astronomical objects.

The Large Binocular Telescope (Hill \& Salinari 1998) has two
Gregorian f/15 identical primaries of 8.4 m diameter, mounted
on the single structure. The LBT
is expected to see first light through the first mirror in
2004 and through the second mirror one year later. First light
instruments include imagers and low resolution spectrographs both
for visible and IR wavelengths (LUCIFER, MODS, LBT Large Binocular Camera).
A HR optical spectrograph is not foreseen for first light, but the
LBT Science Advisory Committee recognized the importance of such
an instrument either as a second generation or as visitor
instrument. 

The parameters space we intend to explore ($R\sim$ 40,000--300,000)
requires highly stabilized and thermalized environment 
and demands an instrument mechanically detached from the
structure. Within the various configurations offered at the LBT
 for a HR spectrograph (Pallavicini~2001), 
the most promising is to position a fiber
feeding unit at each of the two direct Gregorian foci and lead a fiber
bundle out of the LBT structure to the fixed location of the
spectrograph.

Such a configuration allows us to process the light before fiber
feeding via a fore-optics of limited dimensions. This is the case of
the spectropolarimeter PEPSI proposed by Strassmeier~(2001). Each
Gregorian focus feeds a removable polarimeter measuring two
indipendent polarizations, either circular or linear (see
Fig.~\ref{fig:scheme}). Integral light can be measured
bypassing the polarimeters. In either case, four spectra per
echelle order are recorded on the detector (e.g. 2 circular
polarizations + 2 linear polarizations or 2 star spectra + sky +
wavelength reference spectrum). Such an assembly takes full
profit of the double pupil configuration of the LBT and can
provide a high resolution spectro-polarimeter of unprecedented
performances. We emphasize that no high resolution
spectro-polarimeter currently exists on 8--10m class telescopes,
although the possibility of implementing such a polarimeter was
considered for HROS.

Another advantage of the fiber feeding at both Gregorian foci is
the possibility to use the LBT advanced optics. LBT will be
equipped with adaptive secondary mirrors providing AO correction
at the same Gregorian focal stations. Pallavicini (2001) proposed a
very high resolution spectrograph ($R>250,000$) based on the
enhanced throughput in a narrower slit provided by the AO
correction. No spectrograph at present (except UHRF at the 4m
AAT, Diego et al. 1995) is able to reach such a high spectral resolution.

The triple-white-pupil double-echelle $R\varphi \sim$ 80,000
spectrograph we propose combines the performances of both
the above configurations. In the following sections we examine in
some details the characteristics of such a spectrograph.

\section{Preliminary design of a HR and VHR spectrograph for the LBT}

\subsection{Optical design}

We start from the classical double white pupil configuration,
shown in Figures~\ref{fig:white1}~and~\ref{fig:white}, e.g. used in UVES.
Such a configuration forms a first white
pupil at the location of the echelle via an off-axis parabola. The
dispersed light is then focused to an intermediate spectrum close to
the slit location via a second 
passage through the same off-axis parabola and then re-collimated
by another off-axis parabola to reform a second white pupil where
the cross-disperser is located. The light is then focused by a
camera with suitable characteristics onto the CCD.

\begin{figure*}
   \resizebox{\hsize}{!}
   {\includegraphics{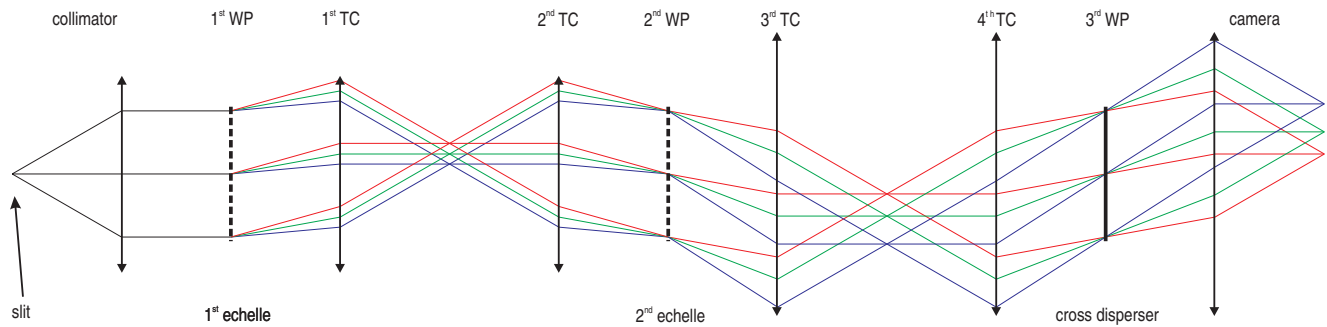}}
   \caption{\textbf{Triple white pupil: schematic design.} A first echelle is
   located in the first white pupil. Two transfer collimators reform a
   second white pupil on the second echelle, and another similar
   system reforms a third white pupil on the cross disperser.
   In our design a parabolic off-axis mirror M1 acts like collimator and
   1$^{st}$ transfer collimator, a second mirror M2 acts like 2$^{nd}$
   and 3$^{rd}$ TC, and a third mirror M3 acts like 4$^{th}$ TC.}
   \label{fig:white2}
\end{figure*}

\begin{figure*}
  \resizebox{\hsize}{!}
   {\includegraphics{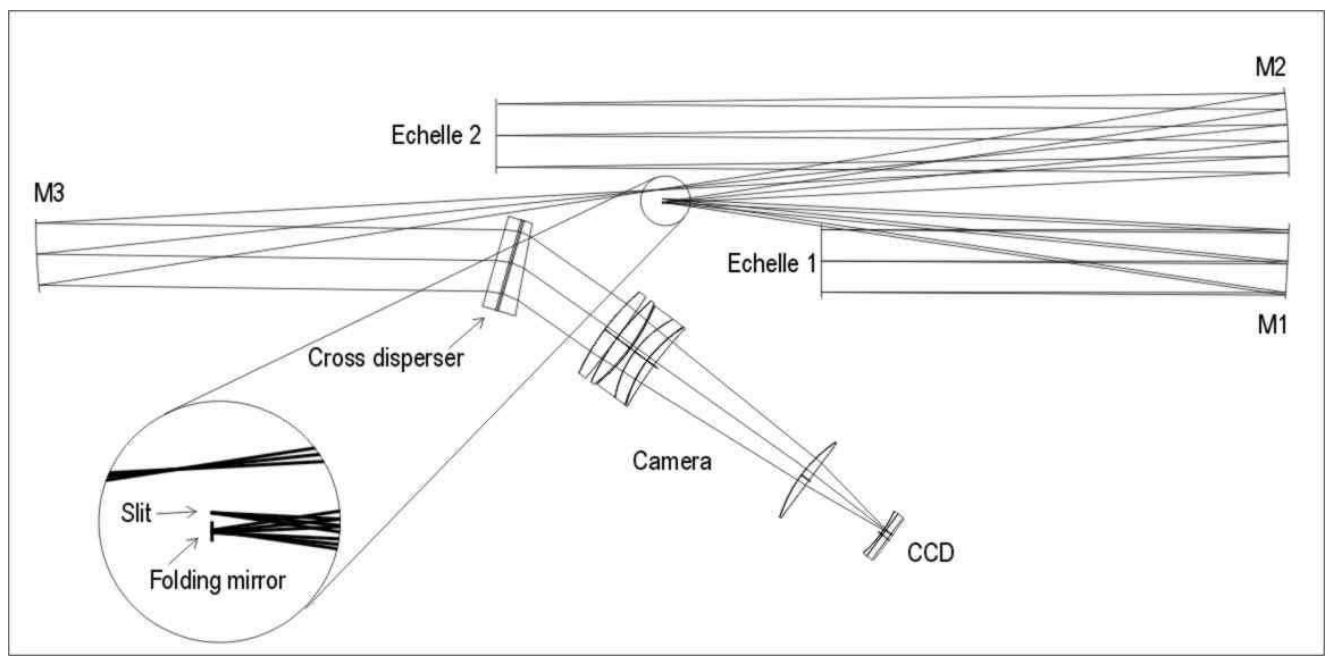}}
  \caption{\textbf{New concept high resolution spectrograph: optical layout.}
  Light from the slit (located near the center of the figure, see
  insert) is collimated from paraboloid M1 onto the
  echelle 1. After diffraction, the beam is refocused again by M1
  near the slit position, where a folding mirror reflects it again towards
  M2. It reforms a white pupil where the echelle 2 is
  located. The beam is diffracted in such a manner as to double its
  angular dispersion. M2 and the other paraboloid M3 reforms another
  white pupil for positioning the cross disperser, and finally the
  beam is focused by a camera (in this design the short
  camera is shown).}
  \label{fig:pepsi1}
\end{figure*}

\begin{figure*}
   \resizebox{\hsize}{!}
   {\includegraphics{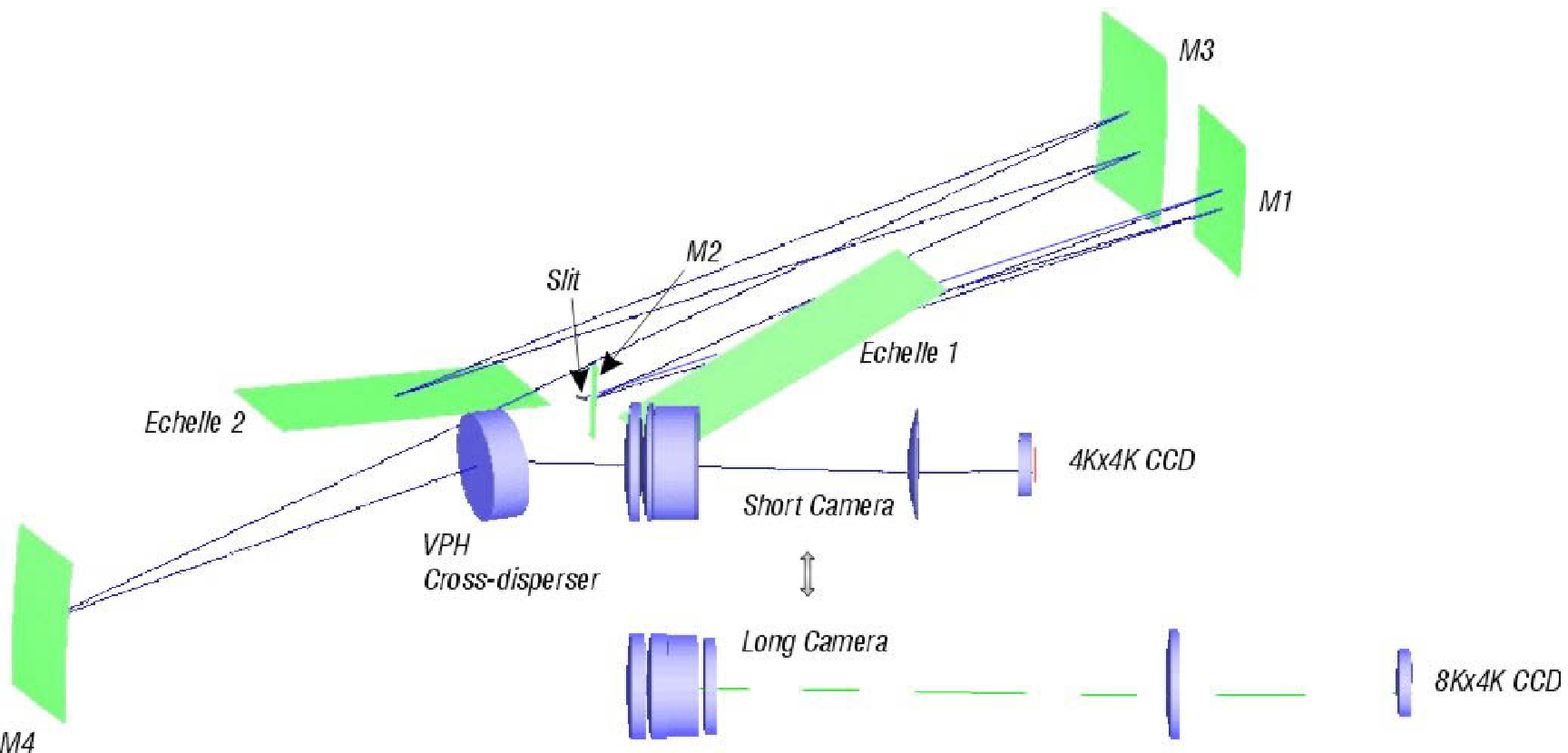}}
   \caption{Spectrograph 3D projection. Note the two tilted R4 echelles
    (the entrance slit is not shown due to its very small size)}
   \label{fig:pepsi2}
\end{figure*}

The first modification occurs at the cross-disperser where
in our design the second R4 echelle is positioned with a small
off-plane angle (see
Figures~\ref{fig:white2},~\ref{fig:pepsi1}~and~\ref{fig:pepsi2}).
After being
diffracted by the second R4 grating the light passes again at the
second parabola and a second intermediate spectrum is reformed
near the slit location. The light is then collimated by a third
off-axis parabola to reform a third white pupil where the
cross-disperser  is located, corresponding to the entrance
pupil of an all-refractive camera.

Due to the high anamorphic effect, in Littrow mode we have to
limit the beam diameter to about 160 mm if we don't want to
vignette the dispersed beam from the first echelle on the second
one: this limits $R\varphi$ to $\sim$65,000. However a slight
off-Littrow angle, the same for both echelles\footnote{From the
grating equations at blaze peak for the two gratings
($m\lambda_{b,1}=2\sigma\sin\delta\cos\theta_1$ and
$m\lambda_{b,2}=2\sigma\sin\delta\cos\theta_2$ where $\theta_i$
are the off-Littrow angles) the blaze wavelengths are the same at
any order only if off-Littrow angles are the same, too.} allows us to
increase $R\varphi$ further. One degree off-littrow allows 
reaching $R\varphi \simeq$ 78,000.

With regard to the echelles, we choose two R4 12.9 gr/mm gratings,
with 1$^{\circ}$ off-Littrow and 0.2$^{\circ}$ off-plane. The
wavelength range from 500 to 1000 nm (\footnote{We are primarily
interested in this spectral range since we want to use the
spectrograph also with adaptive optics; see later.}) is covered in 150 orders,
from the 300$^{th}$  to the 150$^{th}$ order. 
Due to off-Littrow configuration, some additional efficiency losses
reduce total echelles efficiency about 0.30 at blaze peak.

To counterbalance the lower efficiency of the two echelles
configuration as compared
to one echelle, the cross-disperser efficiency must be as  high as
possible. Prisms have too low angular dispersion. Instead of more
common surface-relief reflection gratings, we can use volume phase
holographic (VPH) gratings, which provide measured efficiencies up
to 95\%. However we assume a conservative value about 0.90. 
VPH gratings and their astronomical use are illustrated
by Barden et al.\ (1998).

The two echelles configuration provides an additional interesting
operation mode: a lower resolution mode ($R\varphi \simeq$ 40,000)
obtained by tilting the folding mirror near the slit to exclude
one grating from the optical path. In this mode, after passing on
the first echelle, the beam from M1 is redirected towards M3,
creating a white-pupil directly onto the cross disperser.

\subsection{Fiber feed and adaptive optics}

Since high stability is a primary requirement, 
the spectrograph will be bench mounted and fed with fibers. 
In order to reduce focal
ratio degradation, the fibers cannot be fed directly by the f/15
Gregorian focus. Using  microlenses to modify the focal ratio to
f/5, the scale at the fiber entrance is 204 $\mu$m/arcsec. Two
standard fiber core diameters have been selected, 200 $\mu$m and
50 $\mu$m, corresponding to 1 arcsec and 0.25 arcsec, respectively.
The median seeing at Mt Graham is about 0.7 arcsec, and
200 $\mu$m fibers are well matched to the seeing, carrying most of
the light from the star with little sky contamination. The
smaller fibers are intended to be used with AO correction
for the VHR mode. Since the spectrograph is designed at f/10 the
fibers need also exit microlenses to match the aperture.

When used with the polarimetric units, the spectrograph will receive 2
fibers per Gregorian mirror carrying the pre-processed light. Four
spectra per echelle order have then to be recorded simultaneously
on the CCD. When used instead in integral light, the
light from the star can be directed into two fibers, one for each
telescope. The other two fibers can be used respectively for the sky
and for a wavelength reference spectrum. In this configuration and
assuming the 1 arcsec fibers, the minimum interorder separation
must be about 8 arcsec.
This interorder separation is too small for use of Bowen \& Walraven
image slicers, because these ISs put the slices side by side along the
slit direction. Adaptive Optics is therefore the only
viable solution for a VHR mode in this spectrograph.

\subsection{Cameras, detector and spectral format}

\begin{figure}
  \resizebox{\hsize}{!}
  {\includegraphics[angle=90]{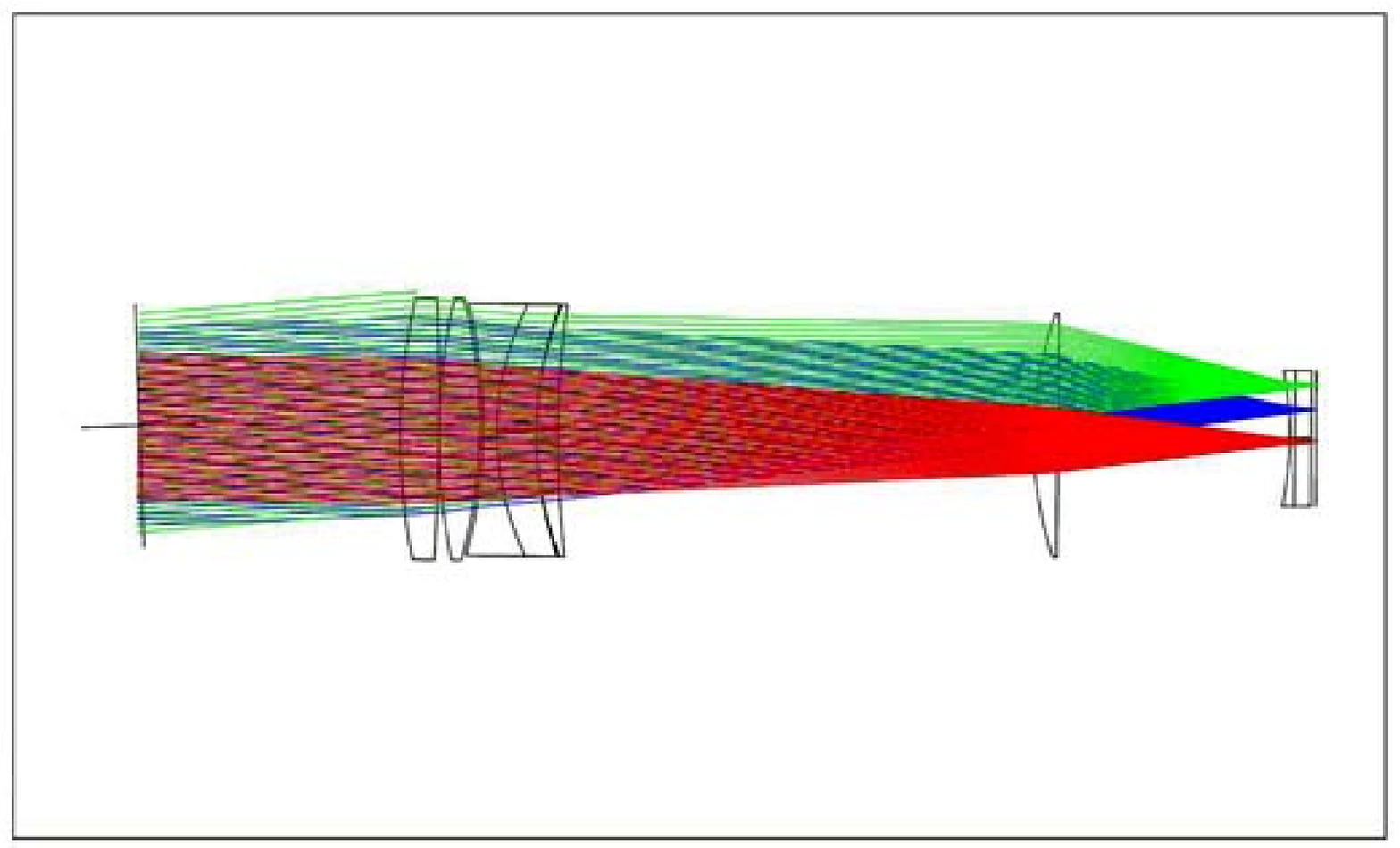}}
  \caption{Short camera: f/3.0, focal length 604 mm. The last lens is
  the dewar window before the CCD. Three different wavelengths
  belonging to the same echelle order are shown. Beam diameter and the
  related focal ratio are wavelength-dependent.}
  \label{fig:camera}
\end{figure}

The design of the camera has to take into account the required
sampling that is linked to the fore-optics characteristics and
ultimately to the fiber core dimensions. It is linked as-well to
the range in spectral resolutions one plans to have in the
spectrograph for each mode, e.g. AO or polarimetric mode. In
the specific case we designed two cameras, one to span resolutions 
from 40,000 to 160,000 and a second one for very
high resolutions (up to 320,000). The short camera has a focal
ratio f/3.0 and a focal length of 604 mm (see
Figure~\ref{fig:camera}). The long camera has a focal ratio f/4.9
and a focal length of 1220 mm.
There is no vignetting with the external focus. The efficiency
of both cameras has been estimated to be about 0.90 
(10 air-glass interfaces).

The evolution of CCD arrays suitable for astronomical
instruments is very fast. Only preliminary considerations
can be drawn about these components at this stage, likely to be
modified during the final design and manufacturing phase. As an
example we have selected a detector with a pixel size 15
$\mu$m, with QE similar to the Marconi deep-depleted fringing-free
CCDs. This chip
has a QE $>$ 40\% in the wavelength range 400-900 nm with a
maximum QE $\geq$  90\% at 600 nm. A 4Kx4K (mosaic) chip of this kind
will be needed together with the short camera and resolution up to
160,000; a mosaic of two 4Kx4K chips will be needed
in the spectrograph detector head for the highest resolution.

The short camera provides a slit sampling of 3 pixels per 0.4
arcsec (corresponding to a maximum resolution of 200,000) at blaze
peak. This is a minimal requirement because a lower sampling (e.g.
2 pixels per 0.4 arcsec) would lead to undersampling in the blue
part of each order, due to the anamorphic magnification (caused by
the two off-Littrow echelles) on the projected slit width. With
the long camera, the slit sampling is 3 pixels per 0.2 arcsec.

\begin{figure}
  \resizebox{\hsize}{!}
  {\includegraphics[width=6cm]{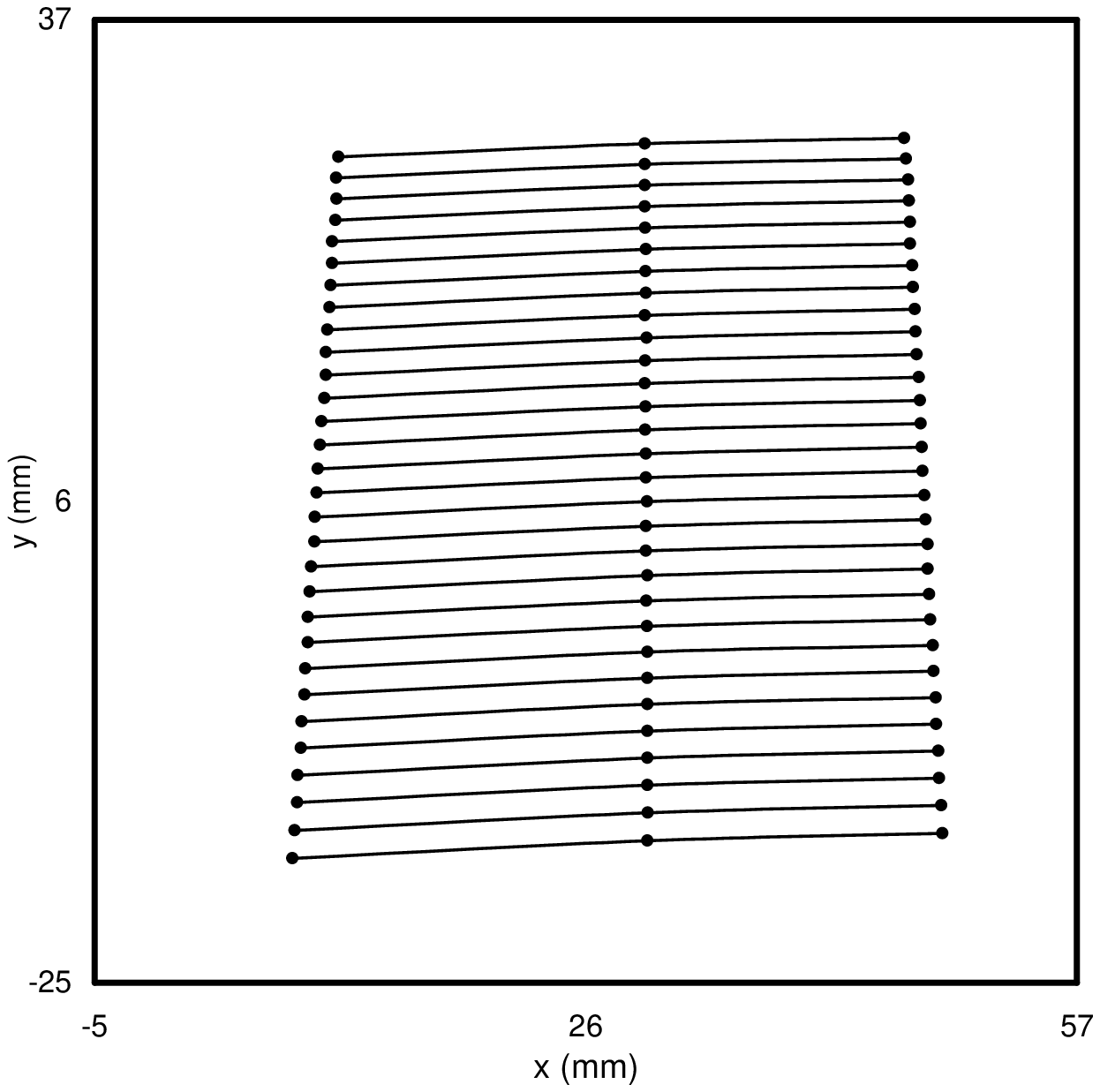}}
  \caption{Typical spectral format with a cross disperser
  and the short
  camera. Interorder separation is about 8 arcsec. The box
  shows the dimensions of a 4Kx4K CCD.}
  \label{fig:specform}
\end{figure}

Assuming 8 arcsec minimum interorder separation, due to the four
spectra per order, from 30 to 45
orders can be accomodated in a 4Kx4K 15 $\mu$m CCD. Thus 4 exposures are
needed to cover the whole spectrum. Each set of orders corresponds to
a different cross disperser, optimized for that wavelength range.
In Figure~\ref{fig:specform} a
typical spectral format with 30 orders is shown.

With the long camera the spectrum must be recorded on a 8Kx4K
CCD mosaic. The use of only one 8Kx4K CCD mosaic with both cameras would be
advisable, using only one half of the CCD area with the short camera.
In this case, a mechanical system is required for switching
between the two cameras.

\subsection{Image Quality}

The whole optical system described above has been optimized taking
simultaneously into account all possible configurations, i.e. the
4 different cross dispersers each coupled with the short or the
long camera. The optimization has been compute for 45 wavelengths
throughout the spectral range of the instrument. Typical RMS spot
radii are about 10-15 $\mu$m  and the
radius corresponding to 80\% encircled energy is less than 15
$\mu$m (1 pixel) for many wavelengths and configurations.
\begin{figure}
  \resizebox{\hsize}{!}
  {\includegraphics{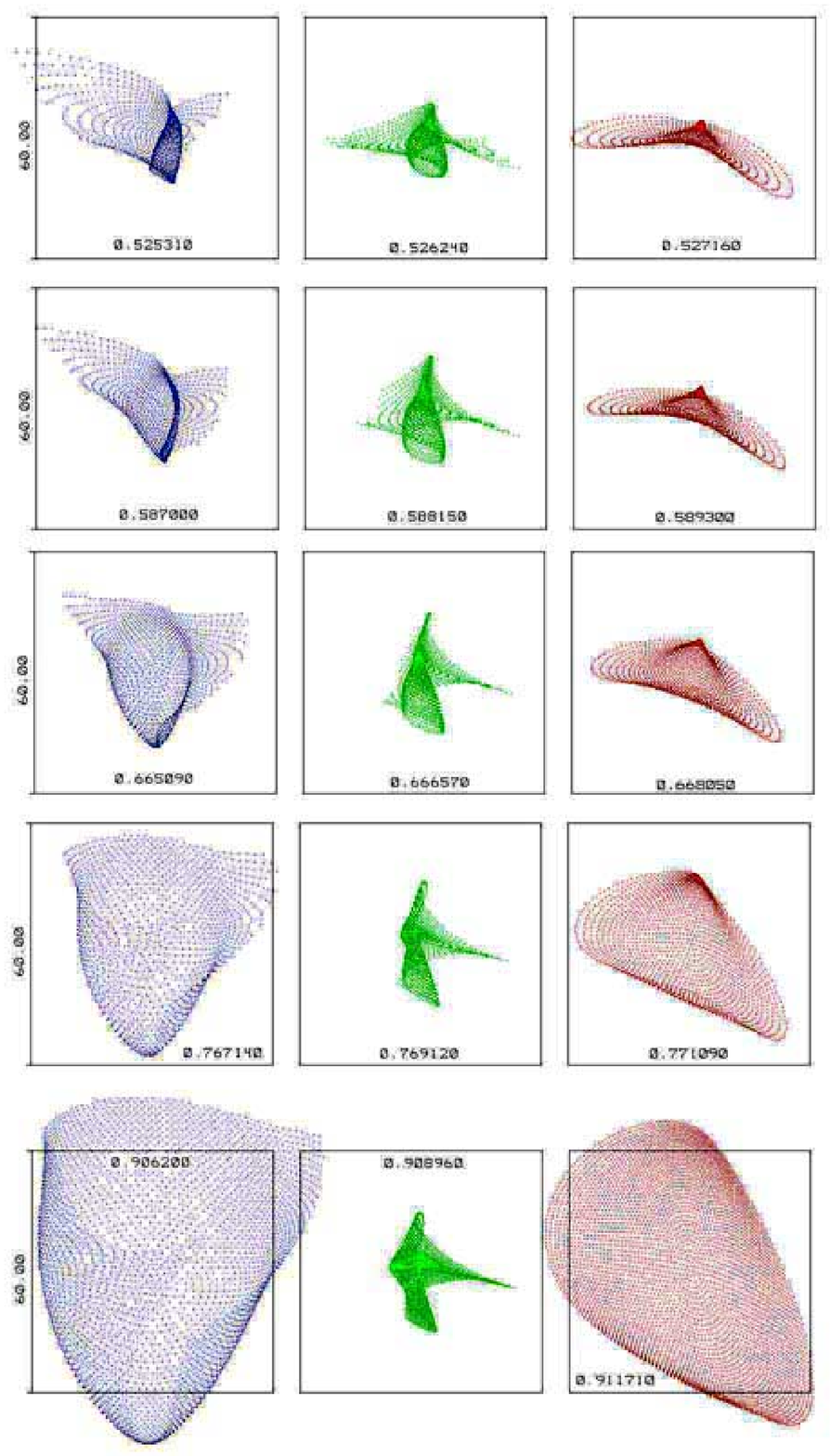}}
  \caption{Matrix spot diagram. Boxs are 4x4 15 $\mu$m pixel
  wide. Wavelengths are measured in micron.}
  \label{fig:spot}
\end{figure}

In Figure~\ref{fig:spot} we show spot diagrams for five different
orders at the blaze peak (second column) and at the borders of the
corresponding orders (first and third columns).
Only at the borders of few orders image quality degrades the maximum
resolution. 

\subsection{Observing modes}

This spectrograph has been designed to provide two main observing
modes: a) High Resolution mode; b) Very High Resolution mode.
The former uses the larger core-radius fiber
bundle and no AO correction. 
The latter instead uses the AO correction and the smaller
core-radius fiber bundle. Each mode has two sub-modes depending on
whether one or two echelles are inserted in the beam. In the former case
(HR mode) two resolutions are available, R=40,000 and R=80,000. In
the latter case (VHR mode) R=160,000 and R=320,000. In 
principle the polarimeter can be used in both the HR and VHR modes,
however the small fraction of polarized ligth in many cases would make its
use in the VHR mode limited to extremely bright objects and hence of
limited interest.

At this level of the design the use of other fiber sizes to
operate the spectrograph at different (or higher) resolutions can
not be excluded. For instance 40 $\mu$m fibers corresponding to
0.2 arcsec on the sky would provide a maximum resolution of about
390,000 in the mode with $R\varphi$ = 78,000. The observing modes
and sub-modes described above are summarized in
Table~\ref{tab:OM}.

\begin{table*}
  \caption{\textbf{Observing modes:} two main modes are avalaible, an
 High Resolution (HR) mode, and a Very High Resolution (VHR) mode. In
 HR mode, with 200 $\mu$m fibers and without AO, resolutions up to
 80,000 are available, while in VHR mode, with 50 $\mu$m fibers and AO,
 resolutions up to 320,000 are available.}
 \centerline{
 \begin{tabular}{@{\extracolsep{\fill}}|c|c|ccccc|c|}  \hline
 Mode & & number of  & fiber size      & AO  & camera & detector & Resolution \\
      & & echelles   & $\mu$m (arcsec) &     &        &     &       \\ \hline\hline
 HR   & (with or w/o & 1 & 200 (1) & OFF & short&4Kx4K& 40,000  \\
      & polarimeter) & 2 & 200 (1) & OFF & short&4kx4K& 80,000  \\ \hline
 VHR  & (no polarimeter) & 1 & 50 (0.25) & ON & short&4Kx4K& 160,000 \\
      &              & 2 & 50 (0.25) & ON & long&8Kx4K& 320,000  \\ \hline
  \end{tabular} }
  \label{tab:OM}
\end{table*}

\subsection{Mechanical Concept}

A possible application of the spectrograph described in these
pages, in fact the one that triggered this study, is its use as
part of the spectropolarimeter PEPSI (Strassmeier 2001, 
Strassmeier et al. 2001) at the LBT.
At the level of the present study the mechanical assembly of the
spectrograph has only been sketched conceptually and we briefly
summarize here the concept. Indeed the final details of such a
design can only be fixed in a phase closer to procurement and
realization.

Due to the requirement of thermal and mechanical stability it is
convenient to have the spectrograph detached from any telescope
movable part. The LBT provides an interesting possible location in
the inner part of the concrete pillar that supports the telescope
main structure. The inner part consists in a cylindrical room of
about 9 m diameter, accessible by stairs and elevator, equipped
with a crane and suitable for a remote instrument location.

At this location the optical components of the spectrograph will
be mounted with adjustable supports on an anti-vibration optical
bench of suitable size. The bench itself will be selected in order
to provide the needed mechanical stability. The upper part of the
bench will be covered with a closure to create a thermal
controlled environment in order to guarantee the needed thermal
stability.

The proposed design makes use of a limited number of movable
parts. Indeed with the exception of the flip mirror allowing to switch
between the polarimetric and the integral ligth modes and of 
a small folding mirror to
toggle between the single-echelle and the double-echelle modes,
the main spectrograph optics (off-axis parabolae and gratings)
have no movable parts. However each of these components has to be
finely adjusted during instrument integration and their supports
designed accordingly.

Two main movable subsystems will instead have to be implemented,
one  before and one after the main spectrograph optics. First the
light that feeds the spectrograph is carried by different fiber
bundles depending on the observational mode. A mechanism to change
the bundles must then be implemented in front of the spectrograph.
Second the full spectral range of the spectrograph requires more
than one cross disperser and this demands an exchange mechanism
for these components after the main spectrograph optics.

Depending on the observing mode the spectrograph needs either the long or
the short camera. These subsystems can be exchanged automatically
by an exchange mechanism or manually at day time. The former
solution allows the rapid change between the corresponding
observing modes but requires high precision mechanics and has a
relevant impact on the cost of the instrument. The latter does not
allow such toggling but is more economic and eventually more
reliable.

\begin{figure*}
  \resizebox{\hsize}{!}
  {\includegraphics{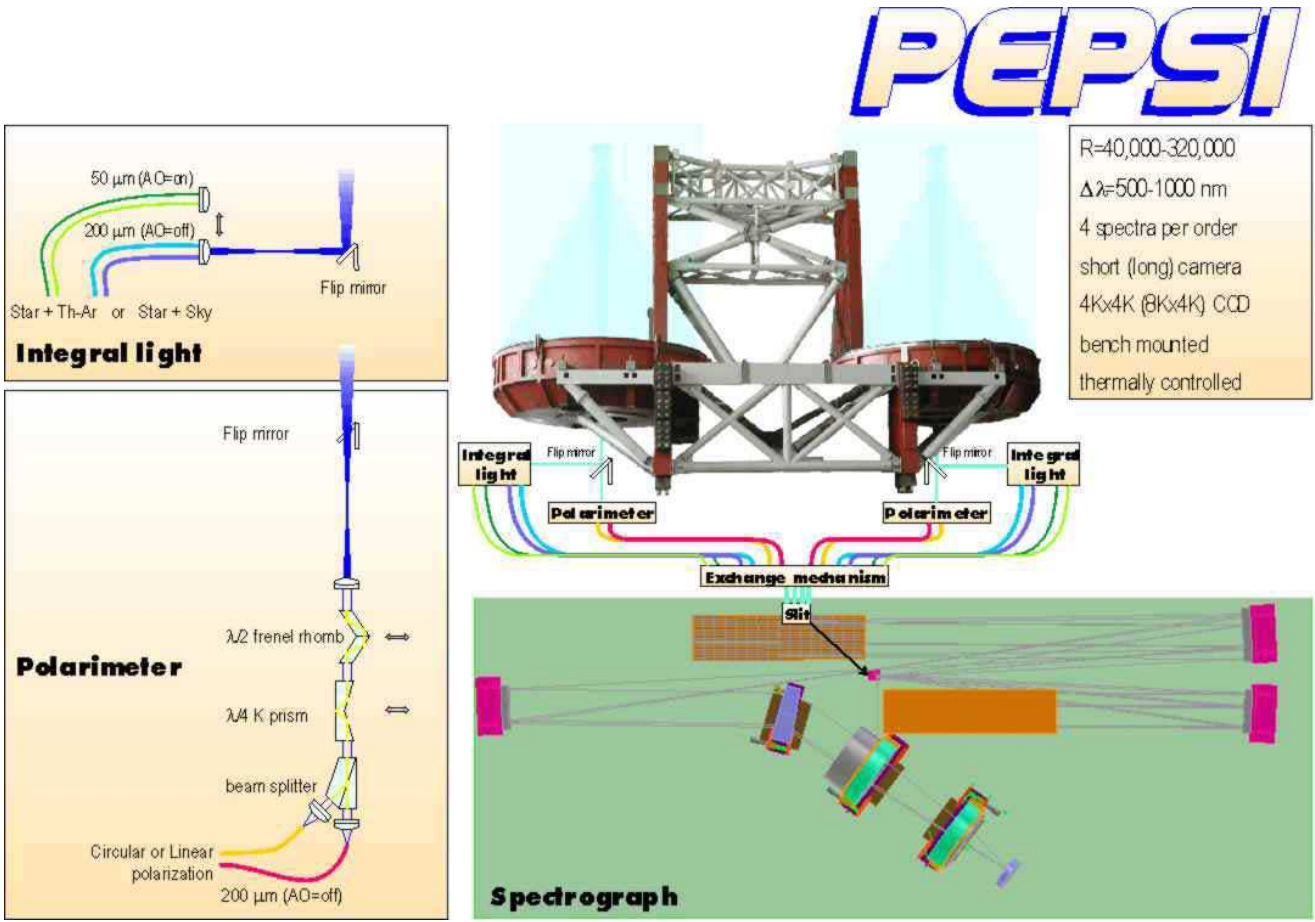}}
  \caption{\label{fig:scheme} Schematic design of the PEPSI instrument at the Large
  Binocular Telescope.}
\end{figure*}

Fig.~\ref{fig:scheme} summarizes the main components of the
instruments in its actual configuration at the LBT.

\subsection{Expected Performances}

The expected performances that can be given at the
current design state are preliminary since they are merely based on
estimates of the efficiency of the components and on theoretical
calculations rather than on measured values.

The design provides a resolution-slit product of  $R
\varphi= 78,000$ at blaze peak. This value is however expected to
vary, due to anamorphic effects, along each order. The
spectrograph has a versatile design that allows us to take advantage of
such unprecedented high $R \varphi$ value in numerous different
configurations.

Throughput has been evaluated under the following assumptions:
a 1\% efficiency loss at
each air-glass interface, 2\% efficiency loss at each (coated) mirror
reflection,  echelles efficiency as given for the two R4 combination
above,  VPH efficiency about 0.90, the QE of a Marconi CCD (basic
midband coated), a telescope efficiency of 0.75 and a fiber efficiency
of 0.80. The calculations have been carried
out at three wavelengths (550, 700 and 850 nm) and the results are
reported in Table~\ref{tab:effi}. The total throughput with no
slit-losses is 0.12 at 550 nm, 0.11 at 700
and 0.07 at 850 nm.

\begin{table}[h!]
\caption{Total throughput for PEPSI.}
\begin{tabular}{|l|ccc|}  \hline
  & \multicolumn{3}{|c|}{Total throughput} \\
 $\lambda$ (nm) & 550 & 700 & 850 \\ \hline
 Optics   & 0.79 & 0.79 & 0.79 \\
 echelles & 0.30 & 0.30 & 0.30 \\
 VPH cross disp. & 0.92 & 0.91 & 0.92 \\
 CCD QE   & 0.92 & 0.87 & 0.57 \\
 telescope & 0.75 & 0.75 & 0.75 \\
 fibers   & 0.80 & 0.80 & 0.80 \\ \hline
 total    & 0.12 & 0.11 & 0.07 \\ \hline
\end{tabular}
\label{tab:effi}
\end{table}

In HR mode at R=60,000 (this resolution, not present in our
design, is chosen for comparison with the other high resolution
spectrographs) the
limiting magnitude in integral light is V=20.2 for S/N=10  per
resolution element in 1 hour (other assumptions are: 1x1 binning,
RON 3 electrons rms per pixel, dark noise 1 e$^-$/pixel/hr, sky
background 22 mag/arcsec$^2$, seeing 0.7 arcsec). 

With the same assumptions and  considering an AO system capable of
concentrating 70\% of the light into a 0.25 arcsec fiber, in VHR mode
at R=320,000 limiting magnitude is V=18.6. 

A comparison of the performances of PEPSI with those of the other HR
spectrographs on 8--10m telescopes is  made in Table~\ref{tab:1}. 

\section{Conclusions}

We have presented in these pages a conceptual design for a versatile
high resolution spectrograph. On the core idea of providing a
resolution-slit product twice that of existing facilities
(80,000 instead of 40,000) with a beam size no larger than in the
latter cases, we designed an instrument capable of
covering resolutions from 40,000 to 320,000 with suitable sampling
and high throughput.

The innovation in the optical concept resides in the use of two R4
echelle grating ``in series'' in a white-pupil configuration and
compensates for the low efficiency 
of this configuration by using Volume Phase Holographic gratings
as cross dispersers.

We have applied the above concept to a fiber-fed bench-mounted 
spectropolarimeter proposed for the Large Binocular Telescope.
In the non-AO mode, light from the two Gregorian foci of the LBT is
preprocessed by two identical polarimetric units (one for each telescope)
which feed by means of a bundle of optical fibers a high-stability thermally
controlled spectrograph mounted on a bench in the telescope pillar.
A cross-dispersed spectrum with four spectra per order, corresponding to
different polarization modes, with a minimum separation of 8 arcsec
between different orders, is recorded on a 4K$\times$4K CCD detector.
The full spectral range between 500 and 1000 nm can be covered in four
exposures. A lower resolution mode (R$\sim$40,000) which makes use of 
only one echelle can be implemented in the same configuration. The
polarimetric units can be either inserted or removed from the optical 
path.
In the latter case, two spectra of the same object (one for each
telescope)  plus sky and a
wavelength reference source are obtained in each exposure.

By using AO and a slit of 0.25 arcsec, a very-high-resolution non-polarimetric 
mode capable of a maximum resolution of $\sim$320,000 can also be implemented 
using a second long camera and a mosaics of two 4k$\times$4K CCD
detectors. This resolution is currently not available at any 8--10m
class telescope. 
By using only one of the two echelles, a maximum resolution of 
$\sim$160,000 can be obtained with the short camera and the same 4k$\times$4K 
CCD detector used in the polarimetric mode. The short and long camera
are interchangeable, either automatically or manually. Different fiber 
sizes (from 200 to 50 $\mu$m) will be  used in the HR (polarimetric) mode
and in the VHR (adaptive optics) mode, respectively.

In its combination of polarimetric capabilities and very-high resolution
modes the proposed instrument will be unique on 8-10m class telescopes
and will allow addressing a large variery of astrophysical problems,
ranging from nearby stars to distant quasars.

\end{document}